\documentclass[sigconf]{acmart}

\usepackage{multirow}
\usepackage{booktabs} 
\setcopyright{rightsretained}

\usepackage{latexsym}
\usepackage{helvet}
\usepackage{courier}
\usepackage{threeparttable}
\usepackage{booktabs, dcolumn}
\usepackage{graphicx}
\usepackage{array,booktabs,ragged2e}
\usepackage[tight,footnotesize]{subfigure}
\usepackage{balance}

\newcommand{\ENEC}{\textit{NamePrism}~}

\newcommand{\ENECE}{\textit{NamePrism}}
\usepackage{color}
\usepackage{amsmath}
\usepackage{amsfonts}
\usepackage{epsfig}
\usepackage{url}
\usepackage{multirow}
\usepackage[ruled,vlined,linesnumbered]{algorithm2e}
\usepackage{setspace}
\usepackage{wrapfig}

\def\BibTeX{{\rm B\kern-.05em{\sc i\kern-.025em b}\kern-.08emT\kern-.1667em\lower.7ex\hbox{E}\kern-.125emX}}

\copyrightyear{2019}
\acmYear{2019}
\setcopyright{acmlicensed}
\acmConference[KDD '19]{KDD '19: 25th ACM SIGKDD Conference on Knowledge Discovery and Data Mining}{August 4 -- 8, 2019}{Anchorage, Alaska USA}

\begin{document}
	
	
	\title{The Secret Lives of Names? Name Embeddings from Social Media}
	
	\author{Junting Ye}
	\affiliation{%
		\institution{Stony Brook University}
		\city{Stony Brook}
		\country{NY}}
	\email{juyye@cs.stonybrook.edu}
	
	\author{Steven Skiena}
	\affiliation{%
		\institution{Stony Brook University}
		\city{Stony Brook}
		\country{NY}}
	\email{skiena@cs.stonybrook.edu}
	
\begin{abstract}
	Your name tells a lot about you: your gender, ethnicity and so on. It has been shown that name embeddings are more effective in representing names than traditional substring features. However, our previous name embedding model is trained on private email data and are not publicly accessible. In this paper, we explore learning name embeddings from public Twitter data. We argue that Twitter embeddings have two key advantages: \textit{(i)} they can and will be publicly released to support research community. \textit{(ii)} even with a smaller training corpus, Twitter embeddings achieve similar performances on multiple tasks comparing to email embeddings.
	
	As a test case to show the power of name embeddings, we investigate the modeling of lifespans. We find it interesting that adding name embeddings can further improve the performances of models using demographic features, which are traditionally used for lifespan modeling. Through residual analysis, we observe that fine-grained groups (potentially reflecting socioeconomic status) are the latent contributing factors encoded in name embeddings. These were previously hidden to demographic models, and may help to enhance the predictive power of a wide class of research studies. 
\end{abstract}

\maketitle
\section{Introduction}
\label{sec:intro}

Your name tells a lot about you. It commonly reveals your gender (male or female) and ethnicity (White, Black, Hispanic, or Asian/Pacific Islander). It can reveal your religion and your country of family origin. It can even inform on your marital status (is it hyphenated?), age (e.g. the generational differences between Fannie and Caitlin), or socioeconomic class (consider Archibald vs. Jethro).

\textit{Name embeddings} are distributed representations which encode the cultural context of name parts (i.e. given name and surname) in 100-dimension vectors learned through an unsupervised technique. It has been shown that name embeddings are more effective representations than substrings on various tasks \cite{ye2017nationality,han2017generating}. Table \ref{tab:example-names-table} presents a representative set of name parts, each with their four nearest neighbors in name embedding space. It is clear that they preserve associations of gender and ethnicity. Unfortunately, previous embeddings were trained on private email data and are not publicly accessible to research community. 

\definecolor{c0}{RGB}{29, 145, 192}
\definecolor{c1}{RGB}{33, 149, 192}
\definecolor{c2}{RGB}{36, 152, 193}
\definecolor{c3}{RGB}{40, 156, 193}
\definecolor{c4}{RGB}{43, 160, 194}
\definecolor{c5}{RGB}{47, 163, 194}
\definecolor{c6}{RGB}{51, 167, 194}
\definecolor{c7}{RGB}{54, 171, 195}
\definecolor{c8}{RGB}{58, 175, 195}
\definecolor{c9}{RGB}{61, 178, 196}
\definecolor{c10}{RGB}{65, 182, 196}
\definecolor{c11}{RGB}{71, 184, 195}
\definecolor{c12}{RGB}{77, 187, 194}
\definecolor{c13}{RGB}{84, 189, 193}
\definecolor{c14}{RGB}{90, 191, 192}
\definecolor{c15}{RGB}{96, 194, 192}
\definecolor{c16}{RGB}{102, 196, 191}
\definecolor{c17}{RGB}{108, 198, 190}
\definecolor{c18}{RGB}{115, 200, 189}
\definecolor{c19}{RGB}{121, 203, 188}
\definecolor{c20}{RGB}{127, 205, 187}
\definecolor{c21}{RGB}{134, 208, 186}
\definecolor{c22}{RGB}{141, 211, 186}
\definecolor{c23}{RGB}{149, 213, 185}
\definecolor{c24}{RGB}{156, 216, 184}
\definecolor{c25}{RGB}{163, 219, 183}
\definecolor{c26}{RGB}{170, 222, 183}
\definecolor{c27}{RGB}{177, 225, 182}
\definecolor{c28}{RGB}{185, 227, 181}
\definecolor{c29}{RGB}{192, 230, 181}
\definecolor{c30}{RGB}{199, 233, 180}
\definecolor{c31}{RGB}{203, 234, 180}
\definecolor{c32}{RGB}{207, 236, 179}
\definecolor{c33}{RGB}{210, 238, 179}
\definecolor{c34}{RGB}{214, 239, 179}
\definecolor{c35}{RGB}{218, 240, 178}
\definecolor{c36}{RGB}{222, 242, 178}
\definecolor{c37}{RGB}{226, 243, 178}
\definecolor{c38}{RGB}{229, 245, 178}
\definecolor{c39}{RGB}{233, 246, 177}
\definecolor{c40}{RGB}{237, 248, 177}
\definecolor{c41}{RGB}{239, 249, 181}
\definecolor{c42}{RGB}{241, 249, 185}
\definecolor{c43}{RGB}{242, 250, 189}
\definecolor{c44}{RGB}{244, 251, 193}
\definecolor{c45}{RGB}{246, 252, 197}
\definecolor{c46}{RGB}{248, 252, 201}
\definecolor{c47}{RGB}{250, 253, 205}
\definecolor{c48}{RGB}{251, 254, 209}
\definecolor{c49}{RGB}{253, 254, 213}
\definecolor{c50}{RGB}{255, 255, 217}
\definecolor{c51}{RGB}{255, 255, 216}
\definecolor{c52}{RGB}{255, 255, 214}
\definecolor{c53}{RGB}{255, 255, 213}
\definecolor{c54}{RGB}{255, 255, 212}
\definecolor{c55}{RGB}{255, 255, 210}
\definecolor{c56}{RGB}{255, 255, 209}
\definecolor{c57}{RGB}{255, 255, 208}
\definecolor{c58}{RGB}{255, 255, 207}
\definecolor{c59}{RGB}{255, 255, 205}
\definecolor{c60}{RGB}{255, 255, 204}
\definecolor{c61}{RGB}{255, 253, 200}
\definecolor{c62}{RGB}{255, 251, 195}
\definecolor{c63}{RGB}{255, 250, 191}
\definecolor{c64}{RGB}{255, 248, 186}
\definecolor{c65}{RGB}{255, 246, 182}
\definecolor{c66}{RGB}{255, 244, 178}
\definecolor{c67}{RGB}{255, 242, 173}
\definecolor{c68}{RGB}{255, 241, 169}
\definecolor{c69}{RGB}{255, 239, 164}
\definecolor{c70}{RGB}{255, 237, 160}
\definecolor{c71}{RGB}{255, 235, 156}
\definecolor{c72}{RGB}{255, 233, 152}
\definecolor{c73}{RGB}{255, 231, 147}
\definecolor{c74}{RGB}{255, 229, 143}
\definecolor{c75}{RGB}{254, 227, 139}
\definecolor{c76}{RGB}{254, 225, 135}
\definecolor{c77}{RGB}{254, 223, 131}
\definecolor{c78}{RGB}{254, 221, 126}
\definecolor{c79}{RGB}{254, 219, 122}
\definecolor{c80}{RGB}{254, 217, 118}
\definecolor{c81}{RGB}{254, 213, 114}
\definecolor{c82}{RGB}{254, 209, 110}
\definecolor{c83}{RGB}{254, 205, 105}
\definecolor{c84}{RGB}{254, 201, 101}
\definecolor{c85}{RGB}{254, 197, 97}
\definecolor{c86}{RGB}{254, 194, 93}
\definecolor{c87}{RGB}{254, 190, 89}
\definecolor{c88}{RGB}{254, 186, 84}
\definecolor{c89}{RGB}{254, 182, 80}
\definecolor{c90}{RGB}{254, 178, 76}
\definecolor{c91}{RGB}{254, 174, 74}
\definecolor{c92}{RGB}{254, 171, 73}
\definecolor{c93}{RGB}{254, 167, 71}
\definecolor{c94}{RGB}{254, 163, 70}
\definecolor{c95}{RGB}{254, 160, 68}
\definecolor{c96}{RGB}{253, 156, 66}
\definecolor{c97}{RGB}{253, 152, 65}
\definecolor{c98}{RGB}{253, 148, 63}
\definecolor{c99}{RGB}{253, 145, 62}
\definecolor{c100}{RGB}{253, 141, 60}

\definecolor{e1}{RGB}{228,26,28}
\definecolor{e2}{RGB}{55,126,184}
\definecolor{e3}{RGB}{77,175,74}
\definecolor{e4}{RGB}{152,78,163}
\definecolor{e5}{RGB}{255,127,0}
\definecolor{e6}{RGB}{166,86,40}
\definecolor{blue}{RGB}{0,0,255}
\definecolor{green}{RGB}{0,128,0}

\definecolor{grey}{RGB}{128,128,128}
\definecolor{k}{RGB}{0,0,0}
\definecolor{g}{RGB}{77,175,74}
\definecolor{r}{RGB}{228,26,28}

\begin{table}[!t]
	\centering
	\begin{tabular}{|p{0.65in}|p{0.45in}p{0.45in}p{0.45in}p{0.45in}|} \hline &&&& \\[-1em]
		Male & 1th NN & 2nd NN & 3rd NN & 4th NN \\ \hline &&&& \\[-1em]
		\textcolor{e1}{Andy} & \textcolor{e1}{Pete} & \textcolor{e1}{Stuart} & \textcolor{e1}{Craig} & \textcolor{e1}{Will}\\
		\textcolor{e2}{Dario} & \textcolor{e2}{Giovanni} & \textcolor{e2}{Luigi} & \textcolor{e2}{Francesco} & \textcolor{e2}{Claudio}\\
		\textcolor{e1}{Hilton} & \textcolor{e1}{Jefferson} & \textcolor{e1}{Maryellen} & \textcolor{e1}{Jayme} & \textcolor{e1}{Brock} \\
		\textcolor{e4}{Lamar} & \textcolor{e4}{Ty} & \textcolor{e4}{Reggie} & \textcolor{e4}{Jada} & \textcolor{e4}{Myles} \\
		\textcolor{e3}{Mohammad} & \textcolor{e3}{Abdul} & \textcolor{e3}{Ahmad} & \textcolor{e3}{Hassan} & \textcolor{e3}{Ahmed} \\
		\textcolor{e2}{Rocco} & \textcolor{e2}{Francesca} & \textcolor{e2}{Carlo} & \textcolor{e2}{Giovanni} & \textcolor{e2}{Luigi} \\  \hline &&&& \\[-1em]
		Female & 1th NN & 2nd NN & 3rd NN & 4th NN  \\ \hline &&&& \\[-1em]
		\textcolor{e4}{Adrienne} & \textcolor{e4}{Aimee} & \textcolor{e4}{Brittany} & \textcolor{e4}{April} & \textcolor{e4}{Kristen} \\
		\textcolor{e3}{Aisha} & \textcolor{e3}{Maryam} & \textcolor{e3}{Fatima} & \textcolor{e3}{Ayesha} & \textcolor{e3}{Fatimah}  \\
		\textcolor{e4}{Brianna} & \textcolor{e4}{Brooke} & \textcolor{e4}{Kayla} & \textcolor{e4}{Kaylee} & \textcolor{e4}{Megan} \\
		\textcolor{e5}{Chan} & \textcolor{e5}{Ka} & \textcolor{e5}{Cherry} & \textcolor{e5}{Yun} & \textcolor{e5}{Sha}  \\
		\textcolor{e4}{Cheyenne} & \textcolor{e4}{Hannah} & \textcolor{e4}{Kayla} & \textcolor{e4}{Madison} & \textcolor{e4}{Kelsey}  \\
		\textcolor{e2}{Gabriella} & \textcolor{e2}{Isabella} & \textcolor{e2}{Dario} & \textcolor{e2}{Cecilia} & \textcolor{e2}{Paola} \\
		\hline
	\end{tabular}
	\caption{Four nearest neighbors of representative names in Twitter embedding space, showing how they preserve gender and ethnicity associations. Notes: {\color{e5} Asian} (Chinese, Korean, Japanese, Vietnamese), {\color{e1} British}, {\color{e2}European} (Spanish, Italian), {\color{e3}Middle Eastern} (Arabic, Hebrew), {\color{e4}North American} (African-American, Native American, Contemporary).}
	\vspace{-.15in}
	\label{tab:example-names-table}
\end{table}%

In this paper, we propose to learn name embeddings from public Twitter data. Our motivation is that name embeddings perform well because of homophily, i.e. the tendency for people to associate with those similar to themselves. These associations are reflected by communication patterns, which explains why large-scale email networks proved so effective at elucidating them. We argue that homophily in communication is universal, and also exists social media \cite{al2012homophily}. Two major properties make Twitter embeddings a better alternative: \textit{(i)} Twitter name embeddings can and will be released to support research community. \textit{(ii)} Twitter embeddings achieve similar performances on gender, ethnicity and nationality identification as Email embeddings, even though the training corpus for Email is two times larger than that for Twitter. We observe that Twitter embeddings have better performances on gender prediction, while Email embeddings achieve higher scores on ethnic predictions.

A second focus of our work is to demonstrate the predictive power of name embeddings on lifespan modeling, where gender, ethnicity and nationality are all contributing features. Average lifespan is one of the most critical measurements associated with quality of life across different demographic groups. Mortality prediction for individuals from available features is the foundation of life insurance industry. Here we demonstrate how an individual's most readily available features (names and corresponding embeddings) can be used to improve the accuracy over comparable demographic models. It is an amazing testament to the power of homophily that contemporary communication patterns can account for mortality in people born over a century ago.

We summarize our primary contributions in this paper as following:

\begin{itemize}
	\item \textit{Twitter name embeddings.} We explore and evaluate nine versions of Twitter name embeddings (see Table \ref{tab:datasets}). We get interesting observations via performance comparisons: \textit{(i)} \textit{Mention} embeddings outperform \textit{Email} embeddings and other Twitter embeddings on gender recognition, indicating stronger gender homophily in Twitter mentions. \textit{(ii)} \textit{Followers} embeddings work better than \textit{Followee}, because ordinary users' followers tend to be family members and/or close friends, while there are more celebrities among followees. The performance is improved after removing celebrities' names from followee lists. \textit{(iii)} \textit{Aggregated*} embeddings perform the best among nine Twitter versions. They have similar vocabulary size and achieve comparable performances on gender, ethnicity and nationality classification as \textit{Email} embeddings. Twitter name embeddings are shared for research community (www.name-prism.com).
	
	\item \textit{Demonstrating the power of name embeddings to improve lifespan modeling.} To demonstrate the power of name embeddings, we train a series of models to predict lifespan as a function of five traditional demographic variables (birth year, state, gender, ethnicity and nationality) and name embedding features. We construct 32 (i.e. $2^5$) different sets of linear regression models containing specific subsets of demographic variables, with and without Twitter/email name embeddings. Incorporating name embeddings in all cases improves the underlying models significantly (p values smaller than 0.01).
	
	\item \textit{Uncovering latent factors encoded in name embeddings.} Implicit feature models, like name embeddings, do not come with natural explanations of exactly what effective properties they are encoding. However, we can gain insight by identifying the names which contribute most strongly to the final model. By conducting residual analysis, we get the most favorable/unfavorable names that increase/decrease lifespan most from the latent factors in name embeddings. We observe that fine-grained groups are the latent contributing factors encoded in name embeddings. For example, our results show that a class-based life-expectancy bias against diminutive names (e.g. Wm, Dan and Guy) as compared to their formal forms (William, Daniel and Guido). In addition, 17 out of 20 most favorable last names have Jewish origins, which agrees with existing observation that Jewish have long average lifespan \cite{abramson2011key}. 
	
\end{itemize}

It is important to note that name embeddings encode homophily as features without explicit labels of gender, ethnicity and nationality. Models which discriminate on such criteria are a growing social concern \cite{o2016weapons}. Name embeddings have the potential to help identify biases, as name embedding-based classifiers \cite{ye2017nationality} are already widely used by over 100 social scientists and economists to study discrimination and homophily \cite{venugopal2017homophily, cochardt2018military, vaanunu2018homophily}. For example, Gornall and Strebulaev find that Asian entrepreneurs received a 6\% higher rate of interested replies than White, after sending 80,000 pitch emails introducing promising but fictitious start-ups to 28,000 venture capitalists \cite{gornall2018gender}. AlShebli et. al. study the effect of diversity on scientific impact, as reflected in citations. They find that ethnic diversity has the strongest correlation with scientific impact \cite{alshebli2018preeminence}. Therefore, we believe a public and sharable name embeddings will help to enhance the predictive power of a wide class of research studies.

\section{Related Work}
\label{sec:related}


\subsection{Names and Mortality}
There have been several previous studies of the impact of names on lifespans. Compared to our work, these have generally been performed on smaller datasets (hundreds or perhaps thousands of individuals), versus the 85 million names in our study. Further, they have generally studied surface features of names as opposed to the latent properties exposed by our name embeddings. In particular, Abel and Kruger \cite{abel2009athletes} observed that
several categories of people whose first name began with `D' appeared to
die earlier than those with other names.
This effect did not show up in a larger-scale study \cite{smith2012people},
and an independent study by Pinzur and Smith
\cite{pinzur2009first}
concludes that first name and life expectancy are not related.

Among athletes, Abel and Krugar \cite{abel2006nicknames} observe that
having nicknames increases longevity.
Shin and Cho \cite{shin2014impact} report that self-reported stress
declines after people legally change their names, demonstrating that
there can be genuine physiological effects associated with undesired names.
Pena's analysis of SSDI data suggests that people with more frequent names
have shorter average and median lifespans.

Nelson and Simmons  \cite{nelson2007moniker} identify several
surprising impacts of names, including that
students whose names begin with C or
D achieve lower GPAs and attend lower-ranked law schools than do students whose
names begin with A or B. 
Jones, et al. \cite{jones2004love} find that
people disproportionately marry others whose first or last name resembles
their own.

\subsection{Gender, Nationality and Ethnicity Detection}
\label{subsec:gne-detection}

Nationality and ethnicity are important demographic categorizations of people, standing in as proxies to represent a range of cultural and historical experiences.
Names are important markers of cultural diversity, and have often served as
the basis of automatic nationality classification for biomedical and sociological research.
In the medical literature, nationality from names has been used as a proxy to reflect genetic differences \cite{burchard2003importance,banda2015characterizing} and public health disparity \cite{barr2014health,quesada2011structural} between groups.
Nationality identification is also important in ads targeting, and
academic studies of political campaigns and social media \cite{chang2010epluribus,appiah2001ethnic}.
Name analysis is often the only practical way to gather ethnicity/nationality
annotations
because of privacy or legal concerns.

Name ethnicity classifiers often make
use of characteristic substrings in names as features \cite{ambekar2009name,chang2010epluribus,treeratpituk2012name}.
Ambekar et al. combine decision tree and Hidden Markov Model to conduct hierarchical classification on a taxonomy with 13 leaf classes \cite{ambekar2009name}.
Treeratpituk et al. utilize both alphabet and phonetics sequences in names to improve performance \cite{treeratpituk2012name}.
Chang et al. use Bayesian methods to infer ethnicity of Facebook users with US census data and study the interactions between ethnic groups \cite{chang2010epluribus}. The linguistics features from users' tweets also reveal their ethnicities \cite{Preotiuc-Pietro18}.
Other relevant efforts are binary ethnicity classifiers on names, e.g. Hispanic vs. Non-Hispanic \cite{buechley1976generally}, Chinese vs. Non-Chinese \cite{coldman1988classification}, South Asian vs. Non-South Asian \cite{harding1999potential}.

The ethnicity/nationality classifier, \ENECE, consists of a 39-class name nationality classifier and a 6-class ethnicity classifier. It uses Naive Bayes model for training and testing on 74 million names labeled with country of residence.
Extensive experiments \cite{ye2017nationality} demonstrate that it
achieves a better classification performance (F1 score) on names drawn from
Wikipedia (0.651) and Email/Twitter (0.795) than competing classifiers
HMM \cite{ambekar2009name}. 
We adopt \ENEC to the ethnicity/nationality classification
for our experiments.

Gender classifiers typically classify names according to statistics on the
ratio of males to females observed in the U.S. Census.
More specifically,
we use data from the 1990 U.S. Census data to label popular first names by gender.
We use these names' labels to approximate that of less common names.
In particular, for a given first name $f$, we find its $k$ nearest neighbors in name embedding space and use a majority vote to decide the gender of $f$.

\section{Name Embeddings}
\label{sec:nameEmbedding}

Distributed representations are feature encodings where objects are represented by points in an abstract $d$-dimensional space, such that similar objects are represented by points close in space. Such representations are a fundamental aspect of Deep Learning \cite{GBC_16}, a recent approach to machine learning which has proven to lead to improved results on many computer vision and natural language processing tasks. Word embeddings are a particularly important type of distributed representation, where each word is denoted by a single point, so that words which play similar roles tend to be represented by nearby points \cite{mikolov2013efficient}.

Inspired by word embeddings, Ye et al. develop name embeddings as a form of distributed representation to capture the semantic meaning of first-name and last-name parts \cite{ye2017nationality}. These new representations were trained on the email contact lists of 57 million people. The use of contact lists is motivated by the principle of homophily: that people generally communicate with people similar to themselves \cite{leskovec2008planetary}. In other words, people disproportionately associate with others of the same gender, ethnicity, nationality, and class. More formally, name embedding algorithm tries to maximize following objective:

\begin{equation}
\label{equ:negative-samplling}
\log \sigma (v_{n}^\top v_{n'}) + \sum_{i=1}^{k} \mathbb{E}_{n_i \sim D} \log \sigma(-v_{n}^\top v_{n_i})
\end{equation}

where $v_n$ is the embedding of name part $n$, and $v_{n'}$ is the embedding of a nearby name part $n'$ that co-occur with $n$ in the same contact list. $n_i$ is a random sample from name part distribution $D$. $\sigma(x)$ is sigmoid function, i.e. $\sigma(x) = 1 / (1 + e^{-x})$.

In a nutshell, the objective aims to maximize the similarities of nearby name part pairs (the first term) and minimize random pair similarities (the second term, i.e. negative sampling). Therefore, the locality properties of name embeddings reflect underlying similarities between name parts, e.g. gender, ethnicity and nationality. 

However, \textit{Email} embeddings and \textit{NamePrism} are corporate property and not shareable \cite{ye2017nationality}. In this section, we discuss how to learn powerful name embeddings from public Twitter data. In particular, we focus on comparing embeddings trained on different user associations from Twitter. We appreciate generous assistance from \textit{NamePrism} team in preparing the experiments.

\subsection{Learning Name Embeddings from Twitter}

We explore the potential of learning name embeddings from Twitter, one of the most popular social media in the world. Its API enables us to access public Tweets and users profiles. In this paper, we are interested in two types of data: \textit{(i)} Tweets containing user associations, including the ones  with user mentions and retweets. \textit{(ii)} follower and followee lists of ordinary users (numbers of followers/followees range between 50 and 500). We assume that follower/followee lists of these users tend to encode more homophily signals than those from celebrities or inactive users.

\subsubsection{Nine Training Corpora}

Nine different Twitter training corpora are prepared to compare strength of different embeddings. Their definitions are as follows. All Names are extracted from Twitter profiles using user IDs. We expect names in the pairs/lists are statistically similar because of homophily. For the convenience of description, let $U_{r}(u_0) = \{u_1, u_2, ..., u_n\}$ be the follower list of user $u_0$, $U_{e}(u_0) = \{u_1, u_2, ..., u_m\}$ be $u_0$'s followee list. 

\begin{itemize}
	\item \textit{Retweet}: Twitter user pairs are extracted from retweets, i.e. $(u_0, u_1)$. $u_0$ posts the retweet and $u_1$ is the original Tweet author.
	\item \textit{Mention}: List of users extracted from Tweets with user mentions, i.e. $(u_0, u_1, ..., u_n)$. $u_0$ is the user posting the Tweet. $u_1$ to $u_n$ are the users mentioned in the Tweet. 
	\item \textit{Follower}: List of users who follow user $u_0$ (i.e. $U_{r}(u_0)$). 
	\item \textit{Followee}: List of users whom user $u_0$ follows (i.e. $U_{e}(u_0)$). 
	\item \textit{Followee*}: We removed celebrities with more than 10,000 followers from followee lists. We assume less homophily between celebrities and fans.
	\item \textit{Friend}: Users whom $u_0$ follows and also who follow $u_0$ (i.e. $U_{f}(u_0) = U_{r}(u_0) \bigcap U_{e}(u_0)$ ).
	\item \textit{NonFriend}: Users who are either followers or followees of $u_0$ but not both (i.e. $U_{f}(u_0) = U_{r}(u_0) \triangle U_{e}(u_0)$ ). 
	\item \textit{Aggregated}: Aggregation of \textit{Retweet}, \textit{Mention}, \textit{Follower} and \textit{Followee}. \textit{Friend} and \textit{NonFriend} are excluded due to redundancy.
	\item \textit{Aggregated*}: An aggregation of \textit{Retweet}, \textit{Mention}, \textit{Follower} and \textit{Followee*}.
\end{itemize}

\begin{figure}[!t]
	\centering
	\includegraphics[width=.48\columnwidth]{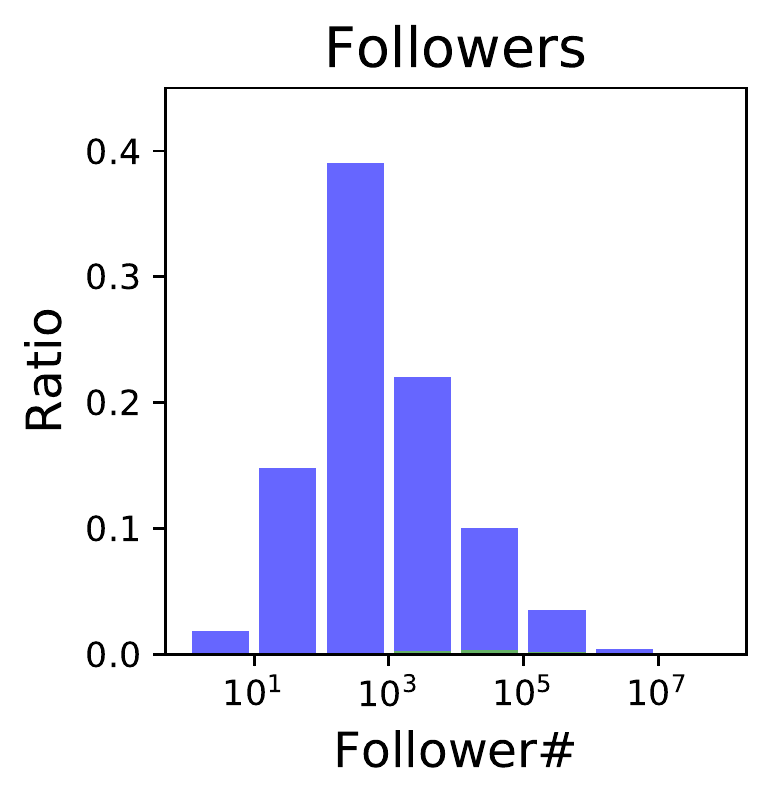} 
	\includegraphics[width=.48\columnwidth]{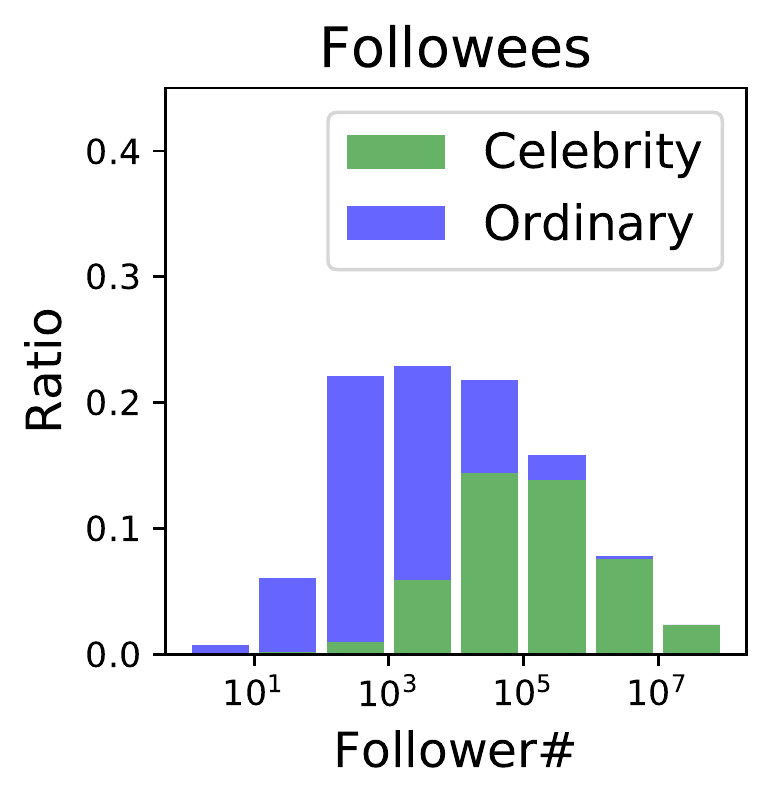}
	\caption{Follower count distributions of \textit{seed users}' followers and followees. We characterize Twitter users with $r$, the ratio of follower over followee count. \textcolor{green}{Celebrity}: $r>10$. \textcolor{blue}{Ordinary}: $r\leq10$. More celebrities among followees. Homophily between fans and celebrities is not as strong as that between families and friends. So \textit{Followee*} removes names of celebrities to strengthen homophily among followee lists.}
	\label{fig:followees_vs_followers}
\end{figure}

\subsubsection{Data Cleaning} Raw data from Twitter can be noisy. Following rules are used to clean data from Twitter API: the first two rules filter out low quality user associations, and the last one normalizes name strings from user profiles.
\begin{itemize} 
	\item \textit{Tweets:} Twitter API provides a small sample of real-time public Tweets\footnote{https://developer.twitter.com/en/docs/tweets/sample-realtime/overview/GET\_statuse\_sample}. On average, we collect about 3.5M Tweets everyday. $\sim$17\% (0.6M) are retweets. $\sim$54\%  (1.9M) contains at least one user mention. Remaining Tweets are filtered out.
	\item  \textit{Users:} In order to get lists of followers and followees, we choose a random set of Twitter users meeting following standards as \textit{seed users}: \textit{(i)} number of followers in range $[50, 500]$. \textit{(ii)} number of followees in range $[50, 500]$. \textit{(iii)} daily average posts less than 10. The motivation is to select Twitter users with enough social links but not celebrities nor social bots \cite{hu2013social}.
	\item  \textit{Names:} Twitter user names can be very noisy, e.g. random strings, misspelled words, emoji and notations. Therefore, we remove special symbols, punctuation and notations in various languages from names. We also filter out names without separators because it is not certain whether they are first or last names. Uncommon names with less than 5 occurrences are also removed.
\end{itemize}

\subsubsection{Followers vs. Followees}
\label{subsubsec:followerfollowee}

After aggregating followers and followees separately, we find these two user groups are fundamentally different. As shown in Figure \ref{fig:followees_vs_followers}, we use a simple but effective way to characterize user, measuring the ratio of follower over followee (referred as $r$). User are assigned label \textit{celebrity} if $r$ is greater than 10, otherwise \textit{ordinary}. 

\begin{table}[!t]
	\centering
	\begin{tabular}{@{}|c|rr|@{}} \hline && \\[-0.95em]
		Embedding & Vocab. Size & Corpus Size \\  \hline && \\[-0.95em]
		\textit{Retweet} & 0.67M & 53.61M \\
		\textit{Mention} & 1.19M & 174.30M \\
		\textit{Follower} & 1.39M & 140.69M \\
		\textit{Followee} & 1.21M & 204.40M \\ 
		\textit{Followee*} & 1.19M & 94.20M  \\ 
		\textit{Friends} & 0.77M & 60.89M \\
		\textit{NonFriends} & 1.30M & 223.25M \\
		\textit{Aggregated} & 3.01M & 573.00M \\ 
		\textit{Aggregated*} &  2.99M & 508.13M  \\ \hline && \\[-0.95em]
		\textit{Email} & 4.10M & 1140.00M \\ \hline 
	\end{tabular}
	\caption{Nine training corpora for Twitter name embeddings. \textit{Email} is baseline corpus. Corpus size of \textit{Followee*} is much smaller than \textit{Followee}, while vocabulary size does not change much.  \textit{Aggregated*} has similar vocabulary and corpus sizes as \textit{Email}.}
	\label{tab:datasets}
	\vspace{-.15in}
\end{table}

As shown in the right, almost half of the followees are celebrities. These celebrities tend to have more than 10,000 followers. We argue that the reason is Twitter allows one-way relation instead of reciprocal relations for \textit{Facebook}. Therefore, ordinary users can follow celebrities they like, as well as their friends and family. As a consequence, these users have more celebrities among their followees and more friends/family among followers. We argue that homophily among friends/family is stronger than that among celebrities and their fans. We will show in Section \textit{Experiments} that performances are improved after removing the celebrities.

\subsubsection{Hyper Parameters}
One of our goals is to compare the performance of Twitter embeddings with email embeddings learned in \cite{ye2017nationality}. Therefore, the same experimental settings are used: skip-gram model with negative sample. Each name part consists of 100 dimensions. The size of moving window for context is 5 and 10 examples for negative sampling. We learn the embeddings for 20 epochs. Strings with less than five occurrences in corpus are ignored.

\subsection{Experiments}
\label{subsec:eval}

\subsubsection{Dataset} 

Two raw datasets have been collected from Twitter for experiments: \textit{(i)} 286 million Tweets are collected from real-time stream sample from Jan. 15 to Mar. 21, 2018. \textit{(ii)} a collection of 922,140 \textit{seed users}' full lists of followers and followees. \textit{Seed users} are collected from real-time Tweet stream. 89 million  unique user profiles are gathered to extract names of the followers and followees. As shown in Table \ref{tab:datasets}, we prepared nine training corpora from this data. \textit{Email} is the dataset used in \cite{ye2017nationality} and it is collected from 57 million email users.

\subsubsection{Performance Comparison}

\begin{figure}[!t]
	\centering
	\includegraphics[width=0.98\columnwidth]{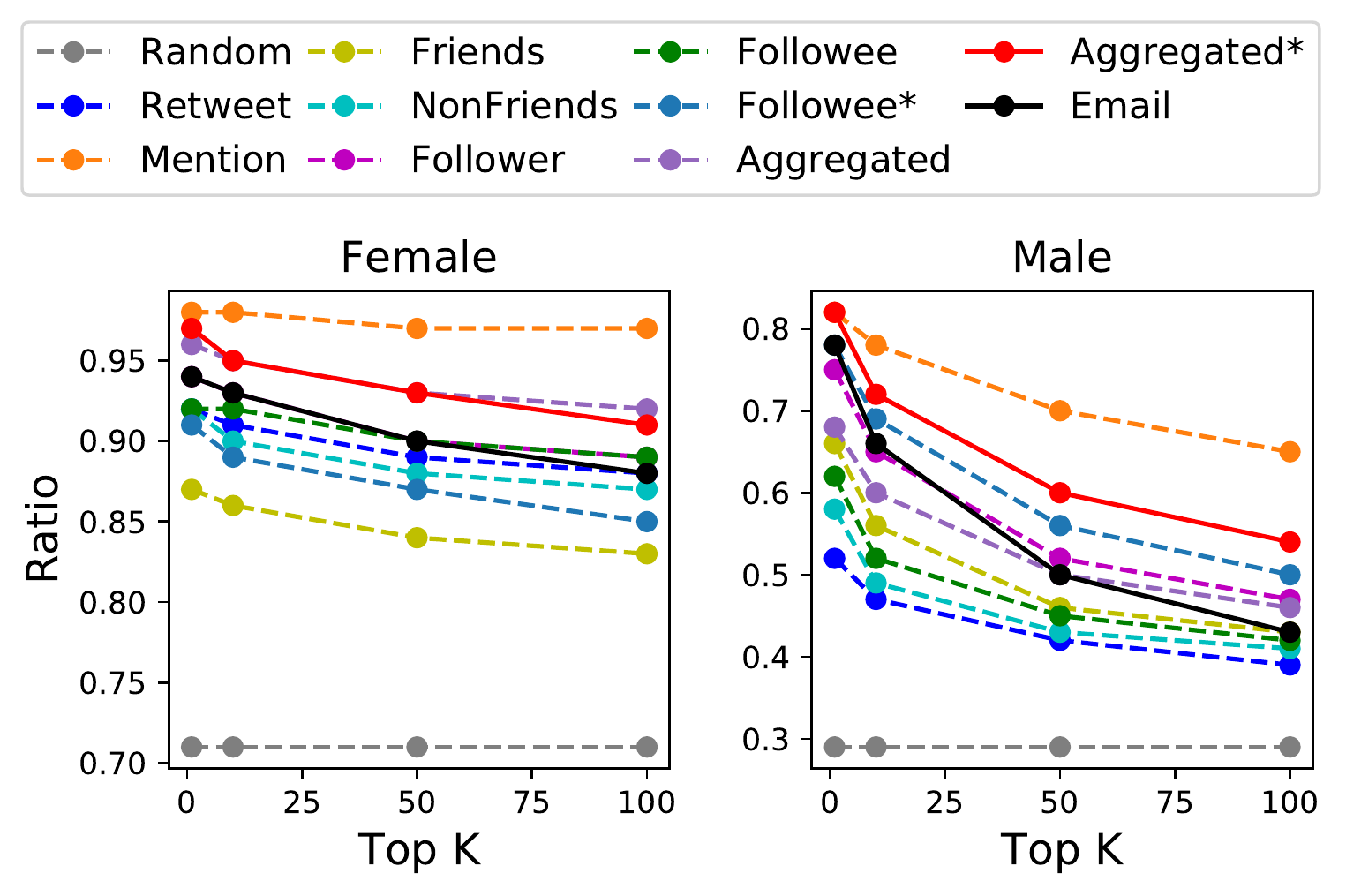}
	\caption{Ratio of same-gender names among top $k$ nearest neighbors ($k \in [1,10,50,100]$) in name embedding spaces.  \textit{Mention} performs the best (avg. on female: 0.94, male: 0.74), reflecting stronger gender homophily in Twitter mentions. \textit{Aggregated} outperforms \textit{Email} on average (female: 0.94 vs. 0.91, male: 0.67 vs. 0.59). \textit{Random} performances are proportional to the ratio of labeled female name count over male.}
	\label{fig:gender}
\end{figure}

Name embeddings prove extremely useful for various tasks because they encode cultural signals of name parts implicitly in the distributed representations. Among the many latent signals, gender, ethnicity and nationalities are major ones that can easily be evaluated. We use ground truth labels from U.S. Census Bureau to measure whether same-gender and same-ethnicity names sit together in embedding space. 74 million name labels from \cite{ye2017nationality} are used to compare classification performances on a 39-leaf nationality taxonomy. 80\% of labels are used for training while 20\% for testing.

\textit{Census 1990} contains ground truth labels of 1,219 male and 4,275 female first names. \textit{Census 2000} provides ethnic distribution of 151,671 last names. We use the names that exist in vocabularies of all embeddings for fair comparison, resulting in 878 male and 3,479 female names for gender evaluation. 58,407 White, 2,519 Black, 4,521 API (Asian and Pacific Islander) and 5,346 Hispanic names are collected for ethnicity in the same manner.

Figure \ref{fig:gender} compares performances on gender. \textit{Mention} embeddings consistently outperform other embeddings by significant margins on both females and males. This suggests that gender bias in Twitter ``mention'' is much stronger than that in ``retweet'' and follower/folloee relations. In other words, Twitter users are more likely to ``@'' others of same gender, who probably share similar interests or opinions. \textit{Aggregated*} has similar vocabulary size as \textit{Email} and achieves better performances than \textit{Email} for both genders. \textit{Followee*} gets a slightly smaller ratio than \textit{followee} on female (avg. 0.88 vs. 0.91) but significantly better on male (avg. 0.65 vs. 0.50).  \textit{Random} embeddings mean that each name part is assigned a random name embedding such that names of each gender uniformly distributed in embedding space. Given a male name, for example, its nearest neighbors have almost the same distribution as the overall gender distribution of the label set. Therefore, we expect performances of male names to be lower than female, because there are far less male name labels (29\% vs. 71\%). We also use similar random embeddings for ethnicity evaluation.

\begin{table}[!t]
	\centering
	\begin{tabular}{@{}|c|cccc|c|@{}} \hline & &&&& \\[-0.95em]
		Embedding &  White & Black & API & Hisp. & Avg.\\  \hline &&&&&  \\[-0.95em]
		\textit{Random} & 0.82 & 0.04 & 0.06 & 0.08 & 0.25 \\ \hline &&&&& \\[-0.95em]
		\textit{Retweet} &  0.92 & 0.20 & 0.57 & 0.64 & 0.58 \\
		\textit{Mention} &  0.93 & 0.22 & 0.61 & 0.71 & 0.62 \\
		\textit{Follower} &  0.94 & 0.31 & 0.77 & 0.86 & 0.72\\
		\textit{Followee} &  0.92 & 0.27 & 0.72 & 0.81& 0.68 \\
		\textit{Followee*} & 0.94 & 0.31 & 0.77 & 0.84 & 0.72 \\
		\textit{Friends} &  0.93 & 0.28 & 0.74 & 0.81 & 0.69 \\
		\textit{NonFriends} &  0.92 & 0.26 & 0.71 & 0.82 & 0.68 \\
		\textit{Aggregated} &  0.93 & 0.32 & 0.76 & 0.83 & 0.71 \\ 
		\textbf{\textit{Aggregated*}} & \textbf{0.94} & \textbf{0.33} & \textbf{0.79} & \textbf{0.86} & \textbf{0.73} \\ \hline &&&&& \\[-0.95em]
		\textbf{\textit{Email}} & \textbf{0.96} & \textbf{0.47} & \textbf{0.83} & \textbf{0.87} & \textbf{0.78} \\ \hline 
	\end{tabular}
	\caption{Ratios of same-ethnicity names among nearest neighbors. \textit{Aggregated*} achieves highest ratios among all Twitter embeddings and gets comparable performance comparing to \textit{Email}. \textit{Follower} outperforms \textit{Followee}, while \textit{Followee*} has the same average ratio as \textit{Follower}, which validates that removing celebrities is effective. (API: Asian and Pacific Islander)}
	\label{tab:eth}
\end{table}

\begin{table*}[t!]
	\centering
	\begin{tabular}{@{}|crcc|crcc|crcc|@{}} \hline &&&&&&& &&&& \\[-1em]
		Nationality & Name\# & \textit{Em.} & \textit{Tw.} & Nationality & Name\# & \textit{Em.} & \textit{Tw.} & Nationality & Name\# & \textit{Em.} & \textit{Twi.} \\ \hline &&&&&&& &&&& \\[-1em]
		CelticEnglish* & 3505K & 0.73 & 0.70 & Muslim & 1475K & 0.74 & 0.73 & Jewish* & 11K & 0.40 & 0.37 \\
		SouthAsian* & 2623K & 0.89 & 0.88 & African & 606K & 0.59 & 0.56 & EastAsian & 6157K & 0.92 & 0.91 \\
		Hispanic & 6892K & 0.91 & 0.89 & Greek* & 259K & 0.89 & 0.87 & Nordic & 195K & 0.73 & 0.70 \\
		Europe & 5371K & 0.84 & 0.81 &  &  &  &  &  &  & & \\\specialrule{1.5pt}{1pt}{1pt} 
		EastEurope* & 65K & 0.49 & 0.49 & Nubian* & 577K & 0.65 & 0.62 & Maghreb* & 47K & 0.15 & 0.14 \\
		SouthKorea* & 68K & 0.86 & 0.83 & Malay & 2596K & 0.86 & 0.84 & Chinese* & 2901K & 0.93 & 0.92 \\
		Portuguese* & 2683K & 0.89 & 0.87 & Turkic & 78K & 0.68 & 0.66 & Pakistanis & 179K & 0.51 & 0.50 \\
		Philippines* & 1137K & 0.72 & 0.69 & Persian* & 423K & 0.66 & 0.64 & Spanish* & 3072K & 0.85 & 0.83 \\
		Scandinavian & 165K & 0.70 & 0.67 & Finland* & 30K & 0.74 & 0.72 & German* & 1278K & 0.74 & 0.70  \\
		WestAfrican* & 315K & 0.56 & 0.54 & Baltics* & 12K & 0.41 & 0.42 & Japan* & 65K & 0.84 & 0.78  \\
		SouthAfrican* & 66K & 0.37 & 0.36 & Russian* & 121K & 0.72 & 0.72 & Arabia* & 172K & 0.51 & 0.51\\
		EastAfrican* & 225K & 0.57 & 0.53 & French* & 2674K & 0.83 & 0.80 & Indochina & 528K & 0.90 & 0.87  \\
		SouthSlavs* & 68K & 0.57 & 0.54 & Italian & 1153K & 0.75 & 0.72 &  &  &  & \\ \specialrule{1.5pt}{1pt}{1pt}
		Cambodia* & 1K & 0.16 & 0.05 & Turkey* & 75K & 0.69 & 0.68 & Sweden* & 74K & 0.61 & 0.58 \\
		Bangladesh* & 78K & 0.58 & 0.56 & Vietnam* & 502K & 0.91 & 0.89 & Thailand* & 18K & 0.59 & 0.67 \\
		Malaysia* & 242K & 0.48 & 0.45 & Pakistan* & 101K & 0.45 & 0.50 & Denmark* & 49K & 0.66 & 0.63 \\
		CentralAsian* & 3K & 0.20 & 0.16 & Italy* & 825K & 0.71 & 0.68 & Romania* & 329K & 0.66 & 0.64 \\
		Indonesia* & 2354K & 0.87 & 0.84 & Norway* & 42K & 0.62 & 0.59 & Myanmar* & 7K & 0.61 & 0.58  \\ \specialrule{1.5pt}{1pt}{1pt}
		\textbf{Weighted Avg.} & --- & \textbf{0.81} & \textbf{0.79} & && & & &  & & \\ \hline
	\end{tabular}
			\caption{Nationality classification performances (f1 scores) of \textit{Email} (\textit{Em.}) and \textit{Twitter (aggregated*, \textit{Tw.})} embeddings on a 39-leaf nationality taxonomy. The taxonomy has three levels, which are separated with bolder lines. `*' marks leaf nationalities. \textit{Weighted Avg.} is count-weighted average F1 score of leaf nationalities. Twitter embeddings achieve comparable performances on nationality classification.}
	\label{tab:largerHierarchyPerf}
		\vspace{-.15in}
\end{table*}

Table \ref{tab:eth} shows the ratios of same-ethnicity last names among their nearest neighbors. It is interesting to see \textit{Mention} gets higher scores than \textit{Retweet}, indicating more ethnic homophily in mentions. One possible explanation is users are more likely to mention or raise attention from their friends while retweeting or quoting from the famous ones. As we have shown in Figure \ref{fig:followees_vs_followers}, there are more celebrities among followees. So \textit{Followee} has lower same-ethnicity ratios than \textit{Followers}. After removing celebrity names, \textit{Followee*} outperforms similarly as \textit{Follower}. The superior performances of \textit{Followee*} over \textit{Followee} on both gender and ethnicities validate less homophily among celebrity-fan pairs and the effectiveness of removing celebrities. Therefore, \textit{Aggregate*} performs best among all Twitter embeddings, after combining training examples from \textit{Followee*} instead of \textit{Followee}. \textit{Email} gets highest ratio among all. \textit{Black} names are harder to classify because they only take up 3.5\% of all labels.

To make a fair comparison on nationality performance, we adopt the same classification method, experiment settings and label data as in \cite{ye2017nationality}. Table \ref{tab:largerHierarchyPerf} shows that \textit{Aggregated*} (\textit{Tw.}) has similar performance as \textit{Email} (\textit{Em.}). For some classes, like Thailand, Baltic and Pakistan, \textit{Aggregated*} outperforms \textit{Email} embeddings. \textit{Email} performs slightly better than \textit{Twitter} w.r.t. weighted average F1 score on 39 leaf classes. We also noticed that the performances are highly dependent on the size of data. For less developed places like Cambodia and countries in central Asian and Maghreb, very limited user associations and labels are collected. Therefore, their F1 scores are much below average performance.

\section{Lifespan Modeling}

\begin{figure}[!t]
	\centering
	\includegraphics[width=6.1cm,height=4.3cm]{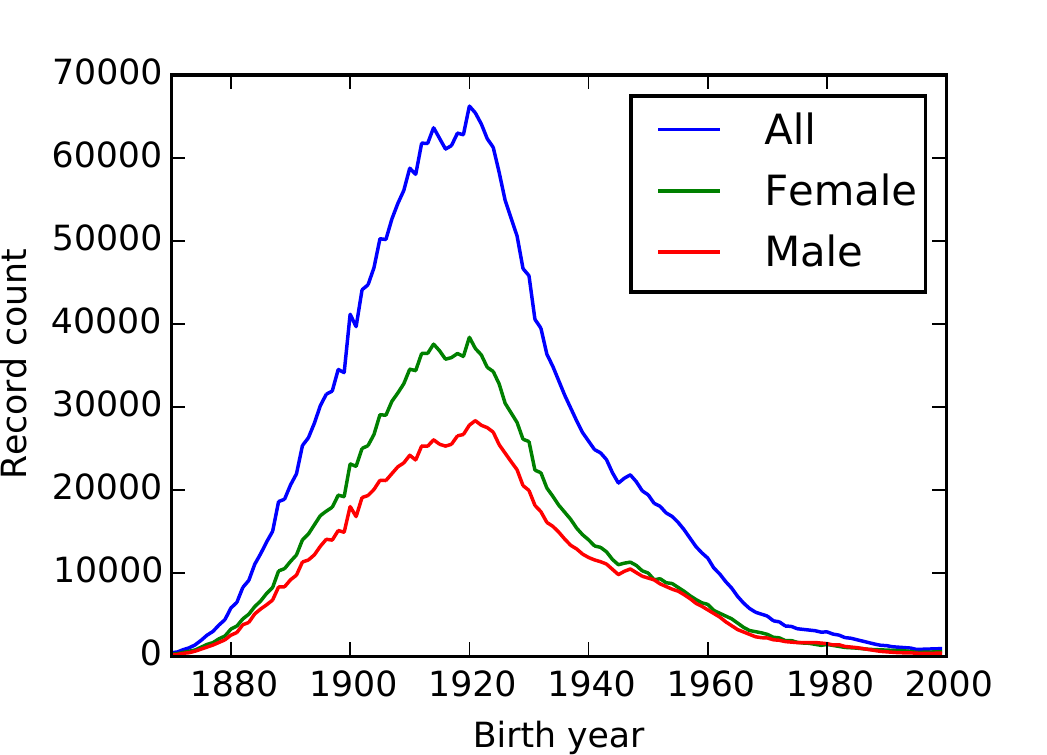}
	\includegraphics[width=6.1cm,height=4.3cm]{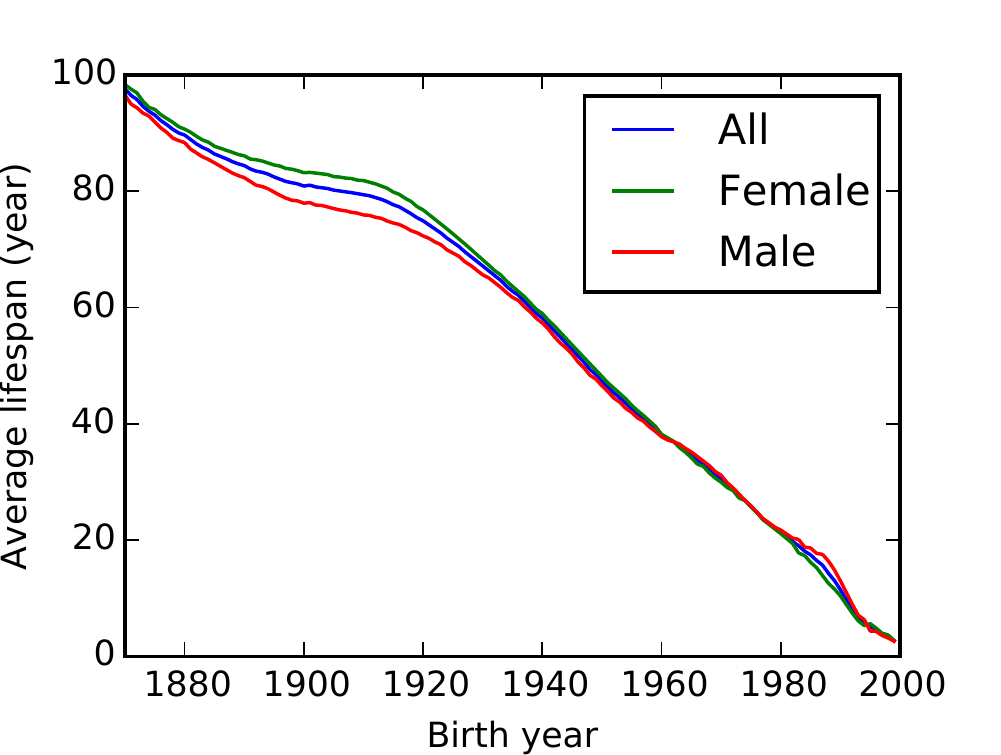}
	\caption{Distributions of SSDI records.  \textit{Top}: the number of records sorted by birth year. Most were born between 1910 and 1930. \textit{Bottom}: the average lifespan by birth year. Survivorship bias causes unusually long lifespan in the beginning, while prematurely deceased ones make decreasing lifepans at the end of the curves.}
	\label{SSDI-demographics}
\end{figure}

The strength of name embeddings lies in the implicit signals encoded in distributed representations. These signals come from concurrences of names, or more accurately, social interactions between individuals (e.g. Tweets). These signals are useful for many downstream tasks. In this section, we demonstrate the power of name embeddings in modeling lifespan, where gender, ethnicity and nationality all are contributing factors.

\subsection{Social Security Death Index Dataset}

\begin{table}[!t]
	\centering
	\begin{tabular}{|@{\hspace{0.03in}}p{0.01in}p{0.01in}p{0.01in}p{0.01in}p{0.05in}|rrrr|@{}} \hline
		\textit{B} & \textit{S} & \textit{G} & \textit{E} & \textit{N} & \textit{NoEbd} & \textit{ShEbd} & \textit{EmEbd} & \textit{TwEbd} \\ \hline &&&&&&&& \\[-0.95em]
		& & & & &  13.418 & 13.423 & 12.781 & 12.747\\\hline  &&&&&&&& \\[-0.95em]
		$\surd$ & & & & & 8.052 & 8.049 & 7.792 & 7.800 \\
		& $\surd$ & &  & & 13.373 & 13.369 & 12.765 & 12.742 \\
		& & $\surd$ & & & 13.150 & 13.135 & 12.768 & 12.745 \\
		& & & $\surd$ & & 13.314 & 13.309 & 12.761 & 12.721 \\
		& & & & $\surd$ & 13.271 & 13.268 & 12.765 & 12.734 \\ \hline &&&&&&&& \\[-0.95em]
		$\surd$ & $\surd$ & $\surd$ & $\surd$ & $\surd$ & 7.775 & 7.777 & 7.739 & 7.744\\ \hline
	\end{tabular}
	\caption{Average Prediction Error (in years) of seven sets of models using different features. The demographic features are: birth year (\textit{B}), state (\textit{S}), gender (\textit{G}), ethnicity (\textit{E}), nationality (\textit{N}). Extra features include: no embedding (\textit{NoEbd}), shuffled embedding (\textit{ShEbd}), Email embedding (\textit{EmEbd}), Twitter embedding (\textit{TwEbd}). Each number (or prediction error) is the average of 20 runs. \textit{Birth year} is the most important feature due to survivorship bias. Using name embeddings improves performance significantly.}
	\label{tab:feat1}
	\vspace{-.15in}
\end{table}

The Social Security Death Index (SSDI) is maintained and distributed by the Social Security Administration to prevent identity fraud associated with using identifiers of deceased individuals. The SSDI has also been employed in hundreds of academic research associated with medical and demographic analysis, such as \cite{backlund1996shape,thompson2013long}. The research applicability of the SSDI compared to other resources has been studied in \cite{rich1994test,williams1992accuracy}.


Each record in the SSDI consists of an individual's full name, their date of birth and death, and their social security number (SSN). The dataset we studied contains 85,822,194 death records. Our analysis was performed on the master file of November 30, 2011\footnote{Dataset available: http://ssdmf.info/}, using a random sample of 2,991,927 records for experiments.

Figure \ref{SSDI-demographics} (Top) presents the number of SSDI records by birth year, further broken down by gender. The peak of the distribution was born between 1910 and 1930. Before this peak, women outnumber men in the database, a consequence of more of them surviving to be issued social security cards. Men and women have represented with equal frequency since approximately 1945. Figure \ref{SSDI-demographics} (Bottom) presents the average lifespan of SSDI records by birth year, further broken down by gender. Survivorship biases account for this strange distribution. The earliest records have an average lifespan above 90, reflecting that they had to live long enough to receive identification numbers (i.e. survivorship bias). The average lifespan has decreased almost linearly since 1940, and equally for woman as for men. We anticipate that these totals will increase and diverge with time, as the distribution moves beyond the prematurely deceased.

\subsection{Demographic Features}
We extracted/inferred following discrete demographic features from each SSDI record, of the type which are traditional for lifespan models. The classifiers for gender, ethnicity and nationality predictions are introduced in Section \textit{Related Work}.

\begin{itemize}
	\item \textit{Birth year}: Birth years are represented by 130 binary features. Each one corresponds to a birth year.
	\item \textit{States}: We infer states using first three digits of SSN. In total, 59 possible binary state/territory features are extracted.
	\item \textit{Gender}: Gender is inferred with a classifier based on U.S. census data. 
	\item \textit{Ethnicity}: We use \textit{NamePrism} to predict ethnicity based on names.
	\item \textit{Nationality}: \textit{NamePrism} is also used to predict nationality based on names.
\end{itemize}

\subsection{Linear Regression Models }

To evaluate the power of name embeddings for predicting lifespan, we build 32 (i.e. $2^5$) sets of models using linear regression. Each set is trained on a particular subset of the 5 demographic features described above. The four models of each set are distinguished by whether they use no embedding features (\textit{NoEbd}), Twitter name embeddings (\textit{TwEbd}), Email name embeddings (\textit{EmEbd}), or a randomly shuffled permutation of Twitter embeddings to add dimensionality without additional information (\textit{ShEbd}), as a control. 

Let $\boldsymbol{X}$ notate the feature vectors, and $\boldsymbol{Y}$ be the ground-truth lifespans. $y_i$ denotes the lifespan of $i_{th}$ record and $\hat{y}_i$ is the predicted lifespan using $i_{th}$ feature vector $x_i$.
Then $|y_i - \hat{y}_i|$ is the error made by prediction (in years). We seek the coefficients $\boldsymbol{w}$ to optimize following loss function:
\vspace{-.05in}

\begin{equation}
\label{equ:ridge}
\min_{\boldsymbol{w}} \sum_i (y_i - \boldsymbol{w}x_i)^T(y_i - \boldsymbol{w}x_i) + \lambda||\boldsymbol{w}||_2^2
\vspace{-.05in}
\end{equation}

Here $\lambda$ is the constant governing the strength of the regularization term, to guard against overfitting. We observed that the performances are not sensitive to $lambda$ and it is empirically assigned 0.003 for all regression models.

\subsection{Performance Analysis}

\begin{figure}[!t]
	\centering
	\hspace{-.35in}
	\includegraphics[width=0.95\columnwidth]{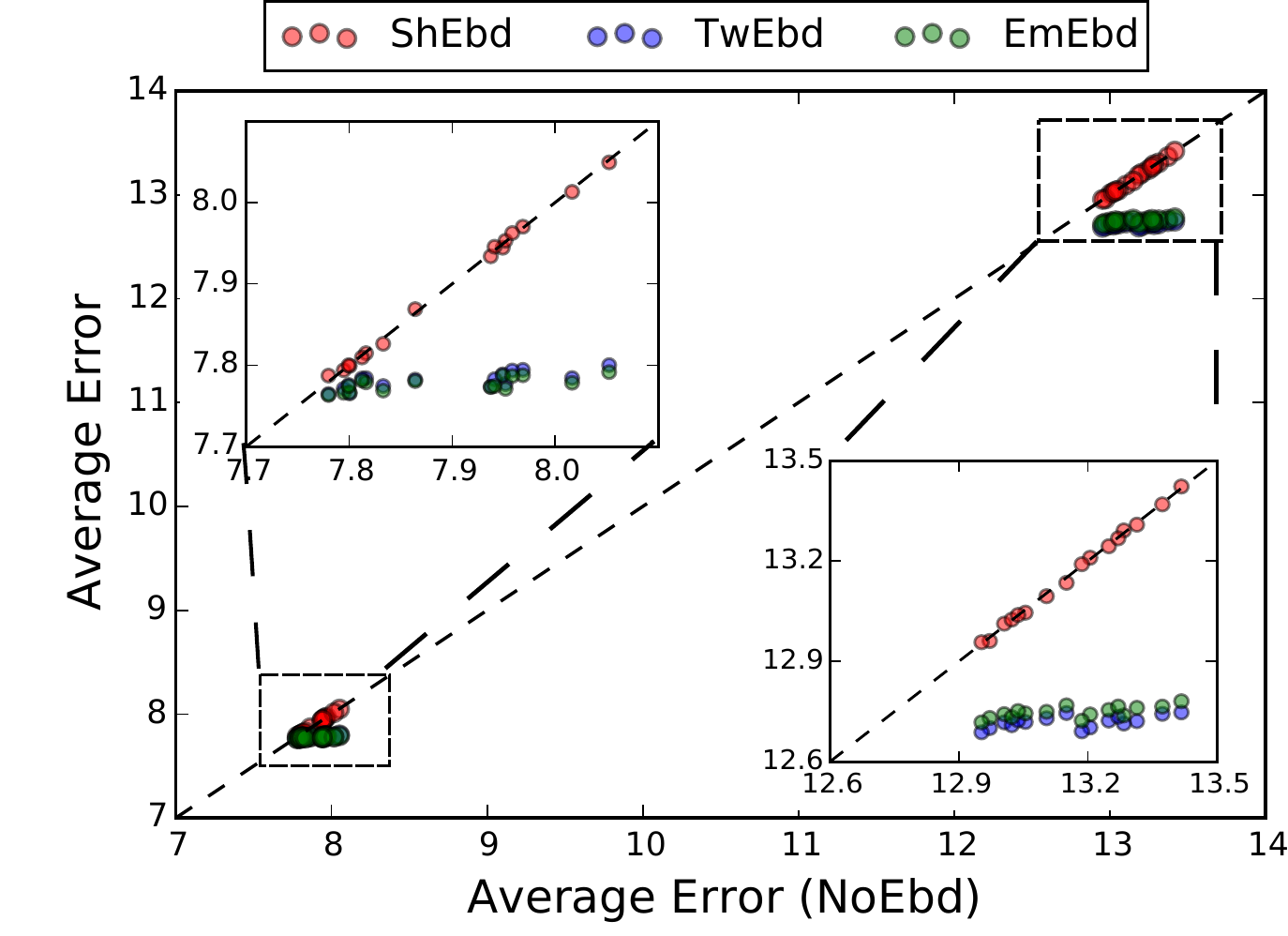}
	\caption{Visualization of prediction errors (in years) of 32 sets of models (two zoom-in inset figures). Using name embeddings (blue and green dots) improve performances significantly (all p values smaller 0.01 under Welch’s t-test). Red dots are on the line, $y=x$, reflecting training without embedding has similar performances as training with shuffled embeddings. Therefore, the gains of name embeddings come from the latent signals captured by name embeddings, instead of the increase of feature dimensions.
	}
	\label{fig:error-visualization}
	\vspace{-.15in}
\end{figure}

We use 90\% records as training data and use the rest as testing data. Table \ref{tab:feat1} presents the average test error of 20 runs after random divisions of training and testing data. Due to survivorship bias, the most powerful single feature is the birth year, which yields an absolute error of 8.052 years. The strength of birth year feature separates models into two groups, with/without birth year (Figure \ref{fig:error-visualization}).

To understand the effect of embeddings on lifespan models, we visualize results of the 32 model sets as points in Figure \ref{fig:error-visualization}. The performance is strikingly linear, and adding name embeddings improves the models of each of the 32 possible variable settings. We also conduct four significance tests using \textit{Welchs t-test} with following null hypotheses: \textit{(i)} the means of \textit{NoEbd} and \textit{ShEbd} are the same; \textit{(ii)} the means of \textit{TwEbd} and \textit{EmEbd} are the same \textit{(iii)} the means of \textit{NoEbd} and \textit{EmEbd} are the same; \textit{(iv)} the means of \textit{NoEbd} and \textit{TwEbd} are the same.  The p values of first two tests under all variable settings are larger than 0.01, and p-values of the last two tests are smaller than 0.01.

The significance tests results show that \textit{(i)} name embeddings further improve the performances of demographic lifespan models; \textit{(ii)} the improvements come from latent signals encoded in name embeddings instead of the increase of feature dimensions. \textit{(iii)} \textit{Twitter} embeddings have similar performances with \textit{Email} embeddings on lifespan modeling.

\subsection{Latent Factors}

Name embeddings are powerful at capturing latent properties of class and cultural group dynamics, but the nature of these properties remains hidden within unlabeled dimensions. This makes it difficult to determine exactly what properties they are keying on for a particular model. To provide some insight into how names affect lifespans, we identified the most favorable/unfavorable first and last names through residual analysis.

\subsubsection{Residual Analysis}
Name embeddings encodes various demographic signals, including gender, ethnicity, nationalities and other latent signals. In our best lifespan model, we combine the explicit demographic features with name embeddings. Therefore, there are redundant signals in the input features. In order to identify latent signals encoded in name embeddings, we conduct residual analysis with following steps: \textit{(i)} train a linear regression model (referred as ``demographic model'') using demographic features and ground-truth lifespan; \textit{(ii)} train linear regression model (referred as ``residual model'')  using name embeddings as features and residuals (i.e. prediction errors) as target values. 

More formally, the demographic model tries to minimize loss function 
$L_d(\boldsymbol{w_{d}} | \boldsymbol{X_d}, \boldsymbol{Y_{ls}})$, where $\boldsymbol{Y_{ls}}$ is the ground truth lifespans and $\boldsymbol{X_d}$ is demographic feature vectors. Let $\boldsymbol{\widehat{Y}_{ls}}$ be the prediction lifespan made by demographic model, i.e. $\boldsymbol{\widehat{Y}_{ls}} = \boldsymbol{w_{d}} \cdot \boldsymbol{X_d}$ (the intercept term is ignored for brevity), then the residual model minimizes $L_r(\boldsymbol{w_{r}} | \boldsymbol{X_e}, \boldsymbol{Y_{ls}} - \boldsymbol{\widehat{Y}_{ls}})$. $\boldsymbol{X_e}$ is name embedding features. $L_d(\cdot | \cdot)$ and $L_r(\cdot | \cdot)$ are in the same form as Equation \ref{equ:ridge}. Finally, we use $\boldsymbol{\widehat{Y}_{r}} = \boldsymbol{w_{r}} \cdot \boldsymbol{X_e}$ to compute gains of name parts. If a name part gets positive gain (i.e. \textit{favorable name}), it means individuals with this name tend to live longer. In the opposite, \textit{unfavorable names} get negative gains.

\begin{table}[!t]
	\centering
	\begin{tabular}{@{}|p{0.3in}rr|p{0.5in}rr|@{}} \hline&&&&& \\[-0.95em]
		Short & Count & Gain & Formal & Count & Gain \\ \hline&&&&& \\[-0.95em]
		Gust & 5429 & -1.965 & Gustav & 8639 & -1.353 \\
		Wm & 6684 & -1.623 & William & 773086 & -0.581 \\
		Gus & 10439 & -1.597 & Angus & 1965 & -0.091 \\
		Hans & 10599 & -1.322 & Johannes & 1153 & -1.053 \\
		Alex & 24520 & -1.297 & Alexander & 34265 & -1.210 \\
		Dan & 12368 & -1.296 & Daniel & 55567 & -0.559 \\
		Guy & 28664 & -1.204 & Guido & 1694 & -0.417 \\
		Effie & 33844 & -1.195 & Euphemia & 753 & 0.073 \\ \hline
		\multicolumn{2}{|c}{Average} & -1.437 & \multicolumn{2}{c}{Average} & -0.649 \\ \hline 
	\end{tabular}
	\caption{8 out of 20 most unfavorable first names are  in diminutive forms. In contrast, their corresponding formal names have larger gains (in years). A systematic study on 155 diminutive/formal name pairs proves bias favoring formal names, suggesting two groups representing different socioeconomic classes.}
	\label{tab:diminutive}
		\vspace{-.15in}
\end{table}

\subsubsection{Diminutive vs Formal First Names}
Among the records with birth year between 1880 to 1910 (less influenced by survivorship bias, see Figure \ref{SSDI-demographics}), 8 out of 20 most unfavorable first names occurring more than 5000 times are in diminutive form (see Table \ref{tab:diminutive}). It is interesting that we find the gains of responding long-version names are significantly larger. We suspect that the distinction captured here is one of socioeconomic class because formal names might be generally expected to appear in official documents more often.

More systematically, we test on 155 pairs of diminutive and formal English names from Wikipedia\footnote{{https://en.wikipedia.org/wiki/Hypocorism}}. It turned out that 114 pairs (74\%) agrees with this observation when using email embeddings, namely names in formal forms get larger gains compared to diminutives. Similarly, 60\% pairs favor formal names using Twitter embeddings. Under the null hypothesis that there is no bias toward short forms or formal ones, email and Twitter embeddings both get p values smaller than 0.01. The null hypothesis is rejected.

\subsubsection{Fine-grained Subgroups:}
Table \ref{tab:origin} shows that, among birth year between 1880 to 1910, 17 out of the 20 most favorable last names are Ashkenazi Jewish. This phenomenon is interesting because Jewish people have long lived in many countries so no single ethnicity or nationality feature could capture this group well. However in name embedding space, similar names have similar representations, because of communication homophily. The observation that these Jewish people had longer life expectancy agrees with the observation made by Institute for Jewish Policy Research\cite{abramson2011key}. It is also interesting to see that popular Scandinavian last names\footnote{https://en.wikipedia.org/wiki/Scandinavian\_family\_name\_\\etymology} all get positive gains.

\begin{table}[!t]
	\centering
	\begin{tabular}{@{}|p{0.5in}rr|p{0.6in}rr|@{}} \hline &&&&& \\[-0.95em]
		\multicolumn{3}{c|}{Jewish} & \multicolumn{3}{c|}{Scandinavian} \\ \hline &&&&& \\[-0.95em]
		Name & Count & Gain & Name & Count & Gain \\ \hline &&&&& \\[-0.95em]
		Katz & 7143 & 1.12 & Svensson & 127 & 1.23 \\ 
		Bernstein & 5122 & 1.11 & Olsson & 647 & 1.16 \\ 
		Shapiro & 6156 & 1.06 & Johansson & 494 & 1.14 \\  
		Solomon & 6232 & 1.03 & Persson & 507 & 1.13 \\ 
		Goldman & 5502 & 1.03 & Karlsson & 129 & 1.13 \\ 
		Levy & 8739 & 1.01 & Nilsson & 682 & 1.01 \\ 
		Feldman & 5031 & 1.01 & Larsson & 207 & 0.90 \\ 
		Friedman & 8865 & 0.97 & Karlsen & 145 & 0.79 \\ 
		Rosenberg & 6448 & 0.95 & Kristiansen & 175 & 0.57 \\ 
		Goldstein & 8837 & 0.95 & Andersen & 6196 & 0.43 \\ 
		Cohen & 22273 & 0.92 & Christensen & 10877 & 0.43 \\ 
		Stern & 5537 & 0.89 & Rasmussen & 5813 & 0.33 \\ 
		Greenberg & 6020 & 0.88 & Pedersen & 3782 & 0.33 \\ 
		Goldberg & 9306 & 0.86 & Larsen & 9488 & 0.31 \\ 
		Levine & 8518 & 0.86 & Hansen & 23578 & 0.29 \\ 
		Rosen & 5363 & 0.85 & Nielsen & 7192 & 0.28 \\ 
		Kessler & 5240 & 0.65 & Olsen & 11080 & 0.23 \\ \hline 
	\end{tabular}
	\caption{17 out of 20 most favorable last name with more than 5000 occurrences have Ashkenazi Jewish origin. Most popular Scandinavian last names get positive gains (in years). Both populations have longer lifespans than the average in U.S. Name embeddings are able to capture such fine-grained distinctions between groups.}
	\label{tab:origin}
	\vspace{-.15in}
\end{table}

\section{Conclusion}
Name embeddings prove more effective feature representations of names than traditional substrings. However, existing \textit{Email} name embeddings are not publicly accessible. In this paper, we present a new way to learn name embeddings from Twitter. Extensive experiment results show the power of Twitter embeddings on gender, ethnicity, nationality. We release Twitter name embeddings to support research communities (www.name-prism.com).

We also demonstrate that name embeddings can improve the accuracy of lifespan models. Extrapolating from these results, we believe they can be used to strengthen predictive models for related tasks in the social, economic, and medical sciences. This is particularly true in large-scale but data-poor studies, where the name must serve as a proxy for reported gender, ethnicity, or nationality. The exact nature of the hidden factors implicitly encoded within our name embeddings that provide this predictive power is an exciting open question for further research. We presume that this includes subtle class-based distinctions (e.g. socioeconomic status and fine-grained groups) which are hidden by the coarse categorical variables traditionally observed and recorded. 

\begin{acks}
The authors thank the reviewers for their useful comments. This work was partially supported by NSF grant IIS-1546113. Any conclusions expressed in this material are of the authors' and do not necessarily reflect the views, either expressed or implied, of the funding party.
\end{acks}

\balance

\bibliographystyle{ACM-Reference-Format}
\bibliography{twitter_embedding}

\end{document}